\documentclass{article}
\usepackage{amsfonts}
\usepackage{amssymb}
\usepackage{amsmath}
\usepackage{makeidx}

\newtheorem{theorem}{Theorem}

\newtheorem{definition}[theorem]{Definition}

\newtheorem{proposition}[theorem]{Proposition}
\newtheorem{remark}[theorem]{Remark}

\input{tcilatex}
\begin{document}

\title{Distinguished Riemann-Hamilton geometry in the polymomentum
electrodynamics}
\author{Alexandru Oan\u{a} and Mircea Neagu}
\date{}
\maketitle

\begin{abstract}
In this paper we develop the distinguished (d-) Riemannian differential
geometry (in the sense of d-connections, d-torsions, d-curvatures and some
geometrical Maxwell-like and Einstein-like equations) for the polymomentum
Hamiltonian which governs the multi-time electrodynamics.
\end{abstract}

\noindent \textit{2010 Mathematics Subject Classification:} 70S05, 53C07,
53C80.

\noindent \textit{Key words and phrases: }jet polymomentum Hamiltonian of
electrodynamics, Cartan canonical connection, Maxwell-like and Einstein-like
equations.

\section{Introduction}

Let $M^{n}$ be a smooth real manifold of dimension $n$, whose local
coordinates are $x=(x^{i})_{i=\overline{1,n}}$, having the physical meaning
of \textit{\textquotedblright space of events\textquotedblright }. In order
to justify the \textit{\textquotedblright electrodynamics\textquotedblright }%
\ terminology used in this paper, we recall that, in the study of classical
electrodynamics, the Lagrangian function $L:TM\rightarrow \mathbb{R}$ that
governs the movement law of a particule of mass $m\neq 0$ and electric
charge $e$, placed concomitantly into a gravitational field and an
electromagnetic one, is given by%
\begin{equation}
L(x,y)=mc\varphi _{ij}(x)y^{i}y^{j}+{\frac{2e}{m}}A_{i}(x)y^{i}+\mathcal{P}%
(x),  \label{ED-Mir-An}
\end{equation}%
where the semi-Riemannian metric $\varphi _{ij}(x)$ represents the \textit{%
gravitational potentials} of the space $M$, $A_{i}(x)$ are the components of
an 1-form on $M$ representing the \textit{electromagnetic potential}, $%
\mathcal{P}(x)$ is a smooth \textit{potential function} on $M$ and $c$ is
the velocity of light in vacuum. The Lagrange space $L^{n}=(M,L(x,y)),$
where $L$ is given by (\ref{ED-Mir-An}), is known in the literature of
specialty as the \textit{autonomous Lagrange space of electrodynamics.} A
deep geometrical study of the Lagrange space $L^{n}$ is now completely done
by Miron and Anastasiei in the book \cite{Miro+Anas}. More general, we point
out that, in the study of classical \textit{rheonomic} (\textit{%
time-dependent}) electrodynamics, a central role is played by the \textit{%
autonomous time-dependent Lagrangian function of electrodynamics} expressed
by%
\begin{equation}
L(t,x,y)=mc\varphi _{ij}(x)y^{i}y^{j}+{\frac{2e}{m}}A_{i}(t,x)y^{i}+\mathcal{%
P}(t,x),  \label{re}
\end{equation}%
where $L:\mathbb{R}\times TM\rightarrow \mathbb{R}$. Note that the
non-dynamical character (i.e., the independence on the temporal coordinate $%
t $) of the spatial semi-Riemannian metric $\varphi _{ij}(x)$ determines the
usage of the term \textit{\textquotedblright autonomous\textquotedblright }
in the preceding definition.

Let $\left( \mathcal{T}^{m},h_{ab}(t)\right) $ be a \textit{%
\textquotedblright multi-time\textquotedblright } smooth Riemannian manifold
of dimension $m$ (please do not confuse with the mass $m\neq 0$), having the
local coordinates $t=(t^{c})_{c=\overline{1,m}}$, and let $J^{1}(\mathcal{T}%
,M)$ be the $1$-jet space produced by the manifolds $\mathcal{T}$ and $M$.
By a natural extension of the preceding examples of electrodynamics
Lagrangian functions, we can consider the jet multi-time Lagrangian function%
\begin{equation}
L(t^{c},x^{k},x_{c}^{k})=mch^{ab}(t)\varphi _{ij}(x)x_{a}^{i}x_{b}^{j}+{%
\frac{2e}{m}}A_{(i)}^{(a)}(t,x)x_{a}^{i}+\mathcal{P}(t,x),  \label{ext-ML-ED}
\end{equation}%
where $A_{(i)}^{(a)}(t,x)$ is a d-tensor on $J^{1}(\mathcal{T},M)$ and $%
\mathcal{P}(t,x)$ is a smooth function on the product manifold $\mathcal{T}%
\times M$.

\begin{remark}
Throughout this paper, the indices $a$, $b$, $c$, $...$ run from $1$ to $m$,
while the indices $i,$ $j,$ $k,$ $...$ run from $1$ to $n$. The Einstein
convention of summation is also adopted all over this work.
\end{remark}

The pair $\mathcal{ED}ML_{m}^{n}=(J^{1}(\mathcal{T},M),L),$ where $L$ is
given by (\ref{ext-ML-ED}), is called the \textit{autonomous multi-time
Lagrange space of electrodynamics}. The distinguished Riemannian
geometrization of the multi-time Lagrange space $\mathcal{ED}ML_{m}^{n}$ is
now completely developed in the Neagu's works \cite{Neag1} and \cite{Neag2}.

Via the classical Legendre transformation, the jet multi-time Lagrangian
function of electrodynamics (\ref{ext-ML-ED}) leads us to the Hamiltonian
function of polymomenta%
\begin{equation}
H=\frac{1}{4mc}h_{ab}\varphi ^{ij}p_{i}^{a}p_{j}^{b}-{\frac{e}{m^{2}c}}%
h_{ab}\varphi ^{ij}A_{(j)}^{(b)}p_{i}^{a}+\frac{e^{2}}{m^{3}c}\left\Vert
A\right\Vert ^{2}-\mathcal{P},  \label{ext-MH-ED}
\end{equation}%
where $H:J^{1\ast }(\mathcal{T},M)\rightarrow \mathbb{R}$, and%
\begin{equation*}
\left\Vert A\right\Vert ^{2}(t,x)=h_{ab}\varphi
^{ij}A_{(i)}^{(a)}A_{(j)}^{(b)}.
\end{equation*}

\begin{definition}
The pair $\mathcal{ED}MH_{m}^{n}=(J^{1\ast }(\mathcal{T},M),H),$ where $H$
is given by (\ref{ext-MH-ED}), is called the \textbf{autonomous multi-time
Hamilton space of electrodynamics}.
\end{definition}

But, using as a pattern the Miron's geometrical ideas from \cite%
{Miro+Hrim+Shim+Saba}, the distinguished Riemannian geometry for quadratic
Hamiltonians of polymomenta (geometry in the sense of d-connections,
d-torsions, d-curvatures and geometrical Maxwell-like and Einstein-like
equations) is constructed on dual 1-jet spaces in the Oan\u{a}-Neagu's paper 
\cite{Oana+Neag-3}. Consequently, in what follows, we apply the general
geometrical result from \cite{Oana+Neag-3} for the particular Hamiltonian
function of polymomenta (\ref{ext-MH-ED}), which governs the multi-time
electrodynamics.

\section{The geometry of the autonomous multi-time Hamilton space of
electrodynamics $\mathcal{ED}MH_{m}^{n}$}

To initiate our Hamiltonian geometrical development for multi-time
electrodynamics, let us consider on the dual $1$-jet space $E^{\ast
}=J^{1\ast }(\mathcal{T},M)$ the \textit{fundamental vertical metrical
d-tensor}%
\begin{equation*}
\Phi _{(a)(b)}^{(i)(j)}={\dfrac{1}{2}}{\frac{\partial ^{2}H}{\partial
p_{i}^{a}\partial p_{j}^{b}}=h_{ab}^{\ast }(t^{c})\varphi ^{ij}(x^{k}),}
\end{equation*}%
where $h_{ab}^{\ast }(t):=\left( 4mc\right) ^{-1}\cdot h_{ab}(t).$ Let $\chi
_{bc}^{a}(t)$ (respectively $\gamma _{ij}^{k}(x)$) be the Christoffel
symbols of the metric $h_{ab}(t)$ (respectively $\varphi _{ij}(x)$).
Obviously, if $\overset{\ast }{\chi }\;\!_{bc}^{a}$ are the Christoffel
symbols of the Riemannian metric ${h_{ab}^{\ast }(t)}$, then we have $%
\overset{\ast }{\chi }\;\!_{bc}^{a}=\chi _{bc}^{a}$.

Using a general result from the geometrical theory of multi-time Hamilton
spaces (see \cite{Atan-Neag2} and \cite{Oana+Neag-3}), by direct
computations, we find

\begin{theorem}
The pair of local functions $N_{\mathcal{ED}}=\left( \underset{1}{N}\text{{}}%
_{(i)b}^{(a)},\text{ }\underset{2}{N}\text{{}}_{(i)j}^{(a)}\right) $ on the
dual $1$-jet space $E^{\ast },$ which are given by%
\begin{equation*}
\begin{array}{l}
\underset{1}{N}\text{{}}_{(i)b}^{(a)}=\chi _{bf}^{a}p_{i}^{f},\medskip \\ 
\underset{2}{N}\text{{}}_{(i)j}^{(a)}=\gamma _{ij}^{r}\left[ \dfrac{2e}{m}%
A_{\left( r\right) }^{\left( a\right) }-p_{r}^{a}\right] -\dfrac{e}{m}\left[ 
\dfrac{\partial A_{(i)}^{(a)}}{\partial x^{j}}+\dfrac{\partial A_{(j)}^{(a)}%
}{\partial x^{i}}\right] ,%
\end{array}%
\end{equation*}%
represents a nonlinear connection on $E^{\ast }$. This nonlinear connection
is called the \textbf{canonical nonlinear connection of the multi-time
Hamilton space of electrodynamics }$\mathcal{ED}MH_{m}^{n}.$
\end{theorem}

Now, let%
\begin{equation*}
\left\{ \dfrac{\delta }{\delta t^{a}},\dfrac{\delta }{\delta x^{i}},\dfrac{%
\partial }{\partial p_{i}^{a}}\right\} \subset \chi \left( E^{\ast }\right)
,\qquad \left\{ dt^{a},dx^{i},\delta p_{i}^{a}\right\} \subset \chi ^{\ast
}\left( E^{\ast }\right)
\end{equation*}%
be the adapted bases produced by the nonlinear connection $N_{\mathcal{ED}}$%
, where 
\begin{equation}
\begin{array}{c}
\dfrac{\delta }{\delta t^{a}}=\dfrac{\partial }{\partial t^{a}}-\underset{1}{%
N}\underset{}{\overset{\left( f\right) }{_{\left( r\right) a}}}\dfrac{%
\partial }{\partial p_{r}^{f}},\medskip \qquad \dfrac{\delta }{\delta x^{i}}=%
\dfrac{\partial }{\partial x^{i}}-\underset{2}{N}\underset{}{\overset{\left(
f\right) }{_{\left( r\right) i}}}\dfrac{\partial }{\partial p_{r}^{f}}, \\ 
\delta p_{i}^{a}=dp_{i}^{a}+\underset{1}{N}\overset{(a)}{\underset{}{%
_{\left( i\right) f}}}dt^{f}+\underset{2}{N}\overset{(a)}{\underset{}{%
_{\left( i\right) r}}}dx^{r}.%
\end{array}
\label{bazele_expl}
\end{equation}

Working with these adapted bases, by direct computations, we can determine
the adapted components of the \textit{generalized Cartan canonical connection%
} of the space $\mathcal{ED}MH_{m}^{n}$, together with its local d-torsions
and d-curvatures (for details, see the general formulas from \cite%
{Oana+Neag-3}).

\begin{theorem}
\begin{enumerate}
\item[\emph{(1)}] The generalized Cartan canonical linear connection of the
autonomous multi-time Hamilton space of electrodynamics $\mathcal{ED}%
MH_{m}^{n}$ is given by%
\begin{equation*}
C\Gamma (N)=\left( \chi _{bc}^{a},\text{ }A_{jc}^{i},\text{ }H_{jk}^{i},%
\text{ }C_{j\left( c\right) }^{i\left( k\right) }\right) ,
\end{equation*}%
where its adapted components are%
\begin{equation}
H_{ab}^{c}=\chi _{ab}^{c},\quad A_{jc}^{i}=0,\quad H_{jk}^{i}=\gamma
_{jk}^{i},\quad C_{j(c)}^{i(k)}=0.  \label{Cartan-ED}
\end{equation}

\item[\emph{(2)}] The torsion $\mathbb{T}$ of the generalized Cartan
canonical linear connection of the space $\mathcal{ED}MH_{m}^{n}$ is
determined by \textbf{three} effective adapted components:%
\begin{equation}
\begin{array}{l}
\medskip {R_{(r)ab}^{(f)}=\chi _{gab}^{f}p_{r}^{g},} \\ 
\medskip {R_{(r)aj}^{(f)}=-\dfrac{2e}{m}\gamma _{rj}^{s}A_{(s);a}^{(f)}+%
\dfrac{e}{m}}\left[ \dfrac{\partial A_{(r)}^{(f)}}{\partial x^{j}}+\dfrac{%
\partial A_{(j)}^{(f)}}{\partial x^{r}}\right] _{;a}{,} \\ 
{R_{(r)ij}^{(f)}=}\mathfrak{R}{_{rij}^{s}\left[ \dfrac{2e}{m}A_{\left(
s\right) }^{\left( f\right) }-p_{s}^{f}\right] -}\dfrac{e}{m}\left[ \dfrac{%
\partial A_{(i)}^{(f)}}{\partial x^{j}}-\dfrac{\partial A_{(j)}^{(f)}}{%
\partial x^{i}}\right] _{:r}{,}%
\end{array}
\label{Torsion-ED}
\end{equation}%
where $\chi _{dab}^{c}(t)$ (respectively $\mathfrak{R}{_{rij}^{k}}(x)$) are
the classical local curvature tensors of the Riemannian metric $h_{ab}(t)$
(respectively semi-Riemannian metric $\varphi _{ij}(x)$), and
\textquotedblright $_{;a}$\textquotedblright\ and \textquotedblright $_{:k}$%
\textquotedblright\ represent the following \textbf{generalized Levi-Civita
covariant derivatives}:
\end{enumerate}

\begin{itemize}
\item the $\mathcal{T}$-generalized Levi-Civita covariant derivative:%
\begin{eqnarray*}
&&\bigskip T_{cj\left( l\right) \left( f\right) ...;a}^{bi\left( d\right)
\left( r\right) ...}\overset{def}{=}\frac{\partial T_{cj\left( l\right)
\left( f\right) ...}^{bi\left( d\right) \left( r\right) ...}}{\partial t^{a}}%
+T_{cj\left( l\right) \left( f\right) ...}^{gi\left( d\right) \left(
r\right) ...}\chi _{ga}^{b}+T_{cj\left( l\right) \left( f\right)
...}^{bi\left( g\right) \left( r\right) ...}\chi _{ga}^{d}+... \\
&&...-T_{gj\left( l\right) \left( f\right) ...}^{bi\left( d\right) \left(
r\right) ...}\chi _{ca}^{g}-T_{cj\left( l\right) \left( g\right)
...}^{bi\left( d\right) \left( r\right) ...}\chi _{fa}^{g}-...
\end{eqnarray*}

\item the $M$-generalized Levi-Civita covariant derivative:%
\begin{eqnarray*}
&&\bigskip T_{cj\left( l\right) \left( f\right) ...:k}^{bi\left( d\right)
\left( r\right) ...}\overset{def}{=}\frac{\partial T_{cj\left( l\right)
\left( f\right) ...}^{bi\left( d\right) \left( r\right) ...}}{\partial x^{k}}%
+T_{cj\left( l\right) \left( f\right) ...}^{bs\left( d\right) \left(
r\right) ...}{\gamma _{sk}^{i}}+T_{cj\left( l\right) \left( f\right)
...}^{bi\left( d\right) \left( s\right) ...}{\gamma _{sk}^{r}}+... \\
&&...-T_{cs\left( l\right) \left( f\right) ...}^{bi\left( d\right) \left(
r\right) ...}{\gamma _{jk}^{s}}-T_{cj\left( s\right) \left( f\right)
...}^{bi\left( d\right) \left( r\right) ...}{\gamma _{lk}^{s}}-...\text{ .}
\end{eqnarray*}
\end{itemize}

\begin{enumerate}
\item[\emph{(3)}] The curvature $\mathbb{R}$ of the Cartan canonical
connection of the space $\mathcal{ED}MH_{m}^{n}$ is determined by the
following \textbf{four} effective adapted components:%
\begin{equation*}
H_{abc}^{d}=\chi _{abc}^{d},\quad R_{ijk}^{l}=\mathfrak{R}_{ijk}^{l}
\end{equation*}%
and%
\begin{equation*}
-R_{(l)(a)bc}^{(d)(i)}=\delta _{l}^{i}\chi _{abc}^{d},\quad
-R_{(i)(a)jk}^{(d)(l)}=-\delta _{a}^{d}\mathfrak{R}_{ijk}^{l}.
\end{equation*}
\end{enumerate}
\end{theorem}

\section{Electromagnetic-like model on the multi-time Hamilton space of
electrodynamics $\mathcal{ED}MH_{m}^{n}$}

In order to describe our geometrical electromagnetic-like theory (depending
on polymomenta) on the multi-time Hamilton space of electrodynamics $%
\mathcal{ED}MH_{m}^{n}$, we underline that, by a simple direct calculation,
we obtain (see \cite{Oana+Neag-3})

\begin{proposition}
The \textbf{metrical deflection d-tensors} of the space $\mathcal{ED}%
MH_{m}^{n}$ are expressed by the formulas:%
\begin{equation}
\begin{array}{l}
\medskip \Delta _{(a)b}^{(i)}=\left[ {h_{af}^{\ast }}\varphi ^{ir}p_{r}^{f}%
\right] _{/b}=0, \\ 
\medskip \Delta _{(a)j}^{(i)}{=[h_{af}^{\ast }\varphi ^{ir}p_{r}^{f}]_{|j}=}%
\dfrac{e}{4m^{2}c}{h_{af}}\varphi ^{ir}\left[ A_{(r):j}^{(f)}+A_{(j):r}^{(f)}%
\right] , \\ 
\vartheta _{(a)(b)}^{(i)(j)}=[{h_{af}^{\ast }}\varphi
^{ir}p_{r}^{f}]|_{(b)}^{(j)}=\dfrac{1}{4mc}h_{ab}\varphi ^{ij},%
\end{array}
\label{Defl-ED}
\end{equation}%
where \textquotedblright $_{/b}$\textquotedblright , \textquotedblright $%
_{|j}$\textquotedblright\ and \textquotedblright $|_{(j)}^{(b)}$%
\textquotedblright\ are the local covariant derivatives induced by the
generalized Cartan canonical connection $C\Gamma \left( N\right) $ (see \cite%
{Oana+Neag} and \cite{Oana+Neag-3}).
\end{proposition}

Moreover, taking into account some general formulas from \cite{Oana+Neag-3},
we introduce

\begin{definition}
The distinguished $2$-form on $J^{1\ast }\left( \mathcal{T},M\right) ,$
locally defined by 
\begin{equation}
\mathbb{F}=F_{(a)j}^{(i)}\delta p_{i}^{a}\wedge
dx^{j}+f_{(a)(b)}^{(i)(j)}\delta p_{i}^{a}\wedge \delta p_{j}^{b},
\label{electromagnetic-NO}
\end{equation}%
where%
\begin{equation}
\begin{array}{c}
\medskip {F_{(a)j}^{(i)}={\dfrac{1}{2}}\left[ \Delta _{(a)j}^{(i)}-\Delta
_{(a)i}^{(j)}\right] =}\dfrac{e}{8m^{2}c}\cdot \underset{\{i,j\}}{{\mathcal{A%
}}}\left\{ {h_{af}}\varphi ^{ir}\left[ A_{(r):j}^{(f)}+A_{(j):r}^{(f)}\right]
\right\} , \\ 
f_{(a)(b)}^{(i)(j)}{={\dfrac{1}{2}}\left[ \vartheta
_{(a)(b)}^{(i)(j)}-\vartheta _{(a)(b)}^{(j)(i)}\right] =0,}%
\end{array}
\label{ElectroMagn-ED}
\end{equation}%
is called the \textbf{polymomentum electromagnetic field attached to the
multi-time Hamilton space of electrodynamics }$\mathcal{ED}MH_{m}^{n}$.
\end{definition}

Now, particularizing the generalized Maxwell-like equations of the
polymomentum electromagnetic field that govern a general multi-time Hamilton
space $MH_{m}^{n}$, we obtain the main result of the polymomentum
electromagnetism on the space $\mathcal{ED}MH_{m}^{n}$ (for more details,
see \cite{Oana+Neag-3}):

\begin{theorem}
The polymomentum electromagnetic components (\ref{ElectroMagn-ED}) of the
autonomous multi-time Hamilton space of electrodynamics $\mathcal{ED}%
MH_{m}^{n}$ are governed by the following \textbf{geometrical Maxwell-like
equations}: 
\begin{equation}
\left\{ 
\begin{array}{l}
\bigskip {F_{(a)j/b}^{(i)}={\dfrac{e\cdot {h_{af}}}{8m^{2}c}\cdot }}\underset%
{\{i,j\}}{\mathcal{A}}\left\{ {\varphi ^{ir}}\left[ \dfrac{\partial
A_{(r)}^{(f)}}{\partial x^{j}}+\dfrac{\partial A_{(j)}^{(f)}}{\partial x^{r}}%
\right] _{;b}{-2\varphi ^{ir}\gamma _{rj}^{s}A_{(s);b}^{(f)}}\right\} \\ 
\bigskip \underset{\{i,j,k\}}{\dsum }{F_{(a)j|k}^{(i)}=-{\dfrac{h_{af}}{8mc}%
\cdot }\underset{\{i,j,k\}}{\dsum }\left\{ \left[ \varphi ^{sr}\mathfrak{R}{%
_{rjk}^{i}-}\varphi ^{ir}\mathfrak{R}{_{rjk}^{s}}\right] p_{s}^{f}+\dfrac{e}{%
m}\cdot \right. } \\ 
\hspace{20mm}\left. \cdot \varphi ^{ir}\left[ 2\mathfrak{R}{%
_{rjk}^{s}A_{(s)}^{(f)}-}\left( \dfrac{\partial A_{(j)}^{(f)}}{\partial x^{k}%
}-\dfrac{\partial A_{(k)}^{(f)}}{\partial x^{j}}\right) _{:r}\right]
\right\} \bigskip \\ 
\underset{\{i,j,k\}}{\dsum }{F_{(a)j}^{(i)}|_{(c)}^{(k)}={0}},%
\end{array}%
\right.  \label{MaxwellEq-ED}
\end{equation}%
where $\mathcal{A}_{\{i,j\}}$ represents an alternate sum and $%
\sum_{\{i,j,k\}}$ represents a cyclic sum.
\end{theorem}

\section{Gravitational-like geometrical model on the multi-time Hamilton
space of electrodynamics}

To expose our geometrical Hamiltonian polymomentum gravitational theory on
the autonomous multi-time Hamilton space of electrodynamics $\mathcal{ED}%
MH_{m}^{n}$, we recall that the fundamental vertical metrical d-tensor%
\begin{equation*}
\Phi _{(a)(b)}^{(i)(j)}=h_{ab}^{\ast }(t)\varphi ^{ij}(x)
\end{equation*}%
and the canonical nonlinear connection%
\begin{equation*}
N_{\mathcal{ED}}=\left( \underset{1}{N}\text{{}}_{(i)b}^{(a)},\text{ }%
\underset{2}{N}\text{{}}_{(i)j}^{(a)}\right)
\end{equation*}%
of the multi-time Hamilton space $\mathcal{ED}MH_{m}^{n}$ produce a
polymomentum gravitational $h^{\ast }$-potential $\mathbb{G}$ on $E^{\ast
}=J^{1\ast }(\mathcal{T},M)$, locally expressed by%
\begin{equation}
\mathbb{G}=h_{ab}^{\ast }dt^{a}\otimes dt^{b}+\varphi _{ij}dx^{i}\otimes
dx^{j}+h_{ab}^{\ast }\varphi ^{ij}\delta p_{i}^{a}\otimes \delta p_{j}^{b}.
\label{Grav-Potential-ED}
\end{equation}

We postulate that the \textit{geometrical Einstein-like equations}, which
govern the multi-time gravitational $h^{\ast }$-potential $\mathbb{G}$\ of
the multi-time Hamilton space of electrodynamics $\mathcal{ED}MH_{m}^{n}$,\
are the abstract geometrical Einstein equations attached to the Cartan
canonical connection $C\Gamma (N)$\ and to the adapted metric $\mathbb{G}$\
on $E^{\ast }$,\ namely%
\begin{equation}
\text{Ric}(C\Gamma )-{\frac{\text{Sc}(C\Gamma )}{2}}\mathbb{G}=\mathcal{K}%
\mathbb{T},  \label{Einstein-Global-Abstract}
\end{equation}%
where Ric$(C\Gamma )$\ represents the \textit{Ricci tensor} of the Cartan
connection, Sc$(C\Gamma )$\ is the \textit{scalar curvature}, $\mathcal{K}$\
is the \textit{Einstein constant} and $\mathbb{T}$\ is an intrinsic d-tensor
of matter, which is called the \textit{stress-energy d-tensor of polymomenta}%
.

In order to describe the local geometrical Einstein-like equations (together
with their generalized conservation laws) in the adapted basis%
\begin{equation*}
\{X_{A}\}={\left\{ {\frac{\delta }{\delta t^{a}}},{\frac{\delta }{\delta
x^{i}}},{\frac{\partial }{\partial p_{i}^{a}}}\right\} },
\end{equation*}%
let $C\Gamma \left( N\right) =(\chi _{ab}^{c},0,\gamma _{jk}^{i},0)$ be the
generalized Cartan canonical connection of the space $\mathcal{ED}MH_{m}^{n}$%
. Taking into account the expressions of its adapted curvature d-tensors on
the space $\mathcal{ED}MH_{m}^{n}$, we immediately find (see \cite%
{Oana+Neag-3}):

\begin{theorem}
The Ricci tensor \emph{Ric}$(C\Gamma )$ of the autonomous multi-time
Hamilton space of electrodynamics $\mathcal{ED}MH_{m}^{n}$ is characterized
by \textbf{two} effective local Ricci d-tensors:%
\begin{equation*}
\chi _{ab}=\chi _{abf}^{f},\qquad\mathfrak{R}_{ij}=\mathfrak{R}_{ijr}^{r}.
\end{equation*}%
These are exactly the classical Ricci tensors of the Riemannian temporal
metric $h_{ab}(t)$ and the semi-Riemannian spatial metric $\varphi _{ij}(x)$.
\end{theorem}

Consequently, using the notations $\chi =h^{ab}\chi _{ab}$ and $\mathfrak{R}%
=\varphi ^{ij}\mathfrak{R}_{ij}$, we get

\begin{theorem}
The scalar curvature \emph{Sc}$(C\Gamma )$ of the generalized Cartan
connection $C\Gamma $ of the space $\mathcal{ED}MH_{m}^{n}$ has the
expression (for details, see \cite{Oana+Neag-3})%
\begin{equation*}
Sc(C\Gamma )=\left( 4mc\right) \cdot \chi +\mathfrak{R},
\end{equation*}%
where $\chi $ and $\mathfrak{R}$ are the classical scalar curvatures of the
semi-Riemannian metrics $h_{ab}(t)$ and $\varphi _{ij}(x)$.
\end{theorem}

Particularizing the generalized Einstein-like equations and the generalized
conservation laws of an arbitrary multi-time Hamilton space $MH_{m}^{n}$, we
can establish the main result of the generalized polymomentum gravitational
theory on the autonomous multi-time Hamilton space of electrodynamics $%
\mathcal{ED}MH_{m}^{n}$ (for more details, see \cite{Oana+Neag-3}):

\begin{theorem}

\begin{enumerate}
\item[\emph{(1)}] The local \textbf{geometrical Einstein-like equations},
that govern the polymomentum gravitational potential of the space $\mathcal{%
ED}MH_{m}^{n}$, have the form%
\begin{equation}
\left\{ 
\begin{array}{l}
\medskip \chi {_{ab}-{\dfrac{\left( 4mc\right) \cdot \chi +\mathfrak{R}}{8mc}%
}h_{ab}=\mathcal{K}}\mathbb{T}{_{ab}} \\ 
\medskip \mathfrak{R}{_{ij}-{\dfrac{\left( 4mc\right) \cdot \chi +\mathfrak{R%
}}{2}}\varphi _{ij}=\mathcal{K}}\mathbb{T}{_{ij}} \\ 
{-{\dfrac{\left( 4mc\right) \cdot \chi +\mathfrak{R}}{8mc}}h_{ab}\varphi
^{ij}=\mathcal{K}}\mathbb{T}_{(a)(b)}^{(i)(j)},%
\end{array}%
\right.  \label{Einstein1-ED}
\end{equation}%
\begin{equation}
\left\{ 
\begin{array}{lll}
\medskip 0=\mathbb{T}_{ai}, & 0=\mathbb{T}_{ia}, & 0=\mathbb{T}_{(a)b}^{(i)}
\\ 
0=\mathbb{T}_{a(b)}^{\text{ \ }\!(j)}, & 0=\mathbb{T}_{i(b)}^{\;(j)}, & 0=%
\mathbb{T}_{(a)j}^{(i)},%
\end{array}%
\right.  \label{Einstein2-ED}
\end{equation}%
where $\mathbb{T}_{AB},\;A,B\in \left\{ a,i,{\QATOP{(i)}{(a)}}\right\} ,$
are the adapted components of the polymomentum stress-energy d-tensor of
matter $\mathbb{T}$.

\item[\emph{(2)}] The \textbf{polymomentum conservation laws} of the
geometrical Einstein-like equations of the space $\mathcal{ED}MH_{m}^{n}$
are expressed by the formulas%
\begin{equation}
\left\{ 
\begin{array}{l}
\medskip {\left[ \left( 4mc\right) \cdot \chi _{b}^{f}-{\dfrac{\left(
4mc\right) \cdot \chi +\mathfrak{R}}{2}}\delta _{b}^{f}\right] _{/f}=0} \\ 
{\left[ \mathfrak{R}_{j}^{r}-{\dfrac{\left( 4mc\right) \cdot \chi +\mathfrak{%
R}}{2}}\delta _{j}^{r}\right] _{|r}=0},%
\end{array}%
\right.  \label{ConsLaws-ED}
\end{equation}%
where $\chi _{b}^{f}=h^{fd}\chi _{db}$ and $\mathfrak{R}_{j}^{r}=\varphi
^{rs}\mathfrak{R}_{sj}$.
\end{enumerate}
\end{theorem}

\textbf{Open Problem.} There exist real physical interpretations for
previous geometrical polymomentum field-like theories, which to be relevant
for the physical domain of electrodynamics?

Alexandru OAN\u{A} and Mircea NEAGU

University Transilvania of Bra\c{s}ov,

Department of Mathematics--Informatics,

Blvd. Iuliu Maniu, no. 50, Bra\c{s}ov 500091, Romania.

\textit{E-mails:} alexandru.oana@unitbv.ro, mircea.neagu@unitbv.ro

\end{document}